# Surface morphology and thickness variation estimation of zeolites via electron ptychography

Enci Zhang[1,2,3], Zhuoya Dong[1,3], Xubin Han[1,2,3], Jianhua Zhang[1,2], Yanhang Ma[1,3*], Huaidong Jiang[1,2,3*]

**Affiliations**

[1] School of Physical Science and Technology, ShanghaiTech University, Shanghai 201210, P. R. China.

[2] Center for Transformative Science, ShanghaiTech University, Shanghai 201210, P. R. China.

[3] Shanghai Key Laboratory of High-resolution Electron Microscopy, ShanghaiTech University, Shanghai 201210, P. R. China.

**Corresponding Author**

**\*Yanhang Ma:**

- School of Physical Science and Technology
- Shanghai Key Laboratory of High-resolution Electron Microscopy, ShanghaiTech University,
Shanghai 201210, China; orcid.org/0000-0003-4814-3740.

**Email:** mayh2@shanghaitech.edu.cn

**\*Huaidong Jiang**

- School of Physical Science and Technology
- Shanghai Key Laboratory of High-resolution Electron Microscopy, ShanghaiTech University,
- Center for Transformative Science, ShanghaiTech University, Shanghai 201210, China.
Shanghai 201210, China; orcid.org/0000-0002-0895-1690

**Email:** jianghd@shanghaitech.edu.cn

**Abstract**

Zeolites, as representative porous materials, possess intricate three-dimensional frameworks that endow them with high surface areas and remarkable catalytic properties. There are a few factors that give a huge influence on the catalytic properties, including the size and connectivity of these three-dimensional channels and atomic level defects. In additional to that, the surface morphology and thickness variation of zeolites particles are essential to their catalytic performances as well. However, it is a significant challenge to characterize these macroscopic properties using conventional techniques due to zeolites' sensitivity to electron beams. In this study, we introduce surface-adaptive electron ptychography, an advanced approach based on multi-slice electron ptychography, which enables high-precision reconstruction of both local atomic configurations and global structural features in zeolite nanoparticles. By adaptively optimizing probe defocus and slice thickness during the reconstruction process, SAEP successfully resolves surface morphology, thickness variations and atomic structure simultaneously. This integrated framework facilitates a direct and intuitive correlation between zeolite channel structures and particle thickness. Our findings open new pathways for large-scale, comprehensive structure–property analysis of beam-sensitive porous materials, deepening the understanding of their catalytic behavior.

**Introduction**

Porous material systems, with zeolites as a typical representative, possess high specific surface areas due to their complex three-dimensional pore structures(1, 2). These materials play a crucial role in various chemical fields, including catalysis(3-5), adsorption and separation(6-8). The size and connectivity of pores(9), as well as local structure of zeolites(10), such as defects and stacking faults(11), significantly influences their catalytic efficiency. In addition to these microscopic features, macroscopic properties such as surface morphology(12, 13) and thickness variation(14, 15) also have a considerable impact on catalytic performance.

Due to the electron beam sensitivity of zeolites, direct measurement of their local surface morphology and thickness variations remains challenging. Electron energy loss spectroscopy (EELS), one of the most commonly used techniques for thickness measurement(16-18), requires extremely high electron doses to ensure accuracy, making it difficult to preserve the structural integrity of zeolites during analysis. Surface morphology characterization of nanoparticle samples is typically performed using scanning electron microscopy (SEM) (19, 20) or atomic-force microscopy (AFM) (21, 22). However, both of them struggle to precisely locate and analyze nanoscale particles for subsequent atomic-resolution imaging. This separation of measurement techniques not only limits the accurate quantification of the physical properties of zeolite nanoparticles, but also hinders the direct correlation with their local atomic structure, making it difficult to establish a complete and coherent structure–property relationship.

Multi-slice electron ptychography (MSEP), as an phase retrieval iterative algorithm primarily designed for parameter optimization based on four-dimensional scanning transmission electron microscopy (4D-STEM), offers unique advantages in atomic-scale structural characterization, including ultra-high sub-angstrom resolution(23-28), exceptional electron dose efficiency, and the capability to reconstruct three-dimensional structures of bulk samples in only one single collection(29-31). These advantages make MSEP particularly valuable for investigating a wide range of beam sensitive materials, including biological cells(32-35), metal-organic frameworks (MOFs)(36), covalent-organic frameworks (COF)(37) and zeolites(38, 39). In addition to that, MSEP has the unique potential to extract not only the atomic structure, but also a range of non-structural parameters from four-dimensional diffraction datasets that contain sample spectroscopic information, such as zone-axis mis-tilt and uniform thickness estimation(40, 41) through parameter optimization demonstrated in previous studies. However, some key factors influencing zeolites' property such as surface morphology and thickness variation have not been discussed yet.

In this study, a new approach called as surface-adaptive electron ptychography (SAEP), is developed to characterize the macroscopic surface morphology and thickness variations with high accuracy of zeolite particle by adaptively optimizing the defocus value of probe and the slice thickness during MSEP reconstruction. By combining the massive atomic structure obtained from conventional MSEP and the corresponding inferred zeolite channel types with the macroscopic information provided by SAEP (**Fig. 1**C), a direct correlation between the microscopic channel structures and the sample thickness is intuitively established. These findings provide an additional perspective for future large-scale characterization, facilitating a deeper understanding of the relationship between the physical structure and chemical properties of zeolites.

**Results and Discussion**

**Theory of SAEP**

The 4D-STEM data collection procedure is as the same with previous, except the sample surface and thickness is no longer seemed as uniform (**Fig. 1A**). In conventional MSEP, the electron beam probe used for reconstruction is typically a probe with fixed defocus value $d_f$, combined with several coherent or incoherent modes for characterization and the bulk sample is divided into slices with fixed thickness $\Delta z$ for reconstruction (**Fig. 1B**). However, in the SAEP algorithm, surface undulations of the sample are translated into variations in the defocus of the probe as the scan position changes. Accordingly, the thickness of the sample slice applied during reconstruction is adjusted dynamically with the scanning position (**Fig. 1C**). This method is achieved based on the widely applied LSQML algorithm(42), leveraging the gradient descent principle. The loss function of LSQML in real space is shown as:

$$E = \sum_r |\chi_{N,r} - \psi_{N,r}|^2$$

Where $r$ is the scan position, $N$ is for the number of sample slices, $\chi_r$ is updated exit-wave function and $\psi_{N,r}$ represents the exit-wave function before reciprocal space constrain.

The variations of the defocus value can be indirectly obtained by optimizing the probe propagation distance that governs the interaction between the initialized electron probe with a fixed defocus value and the sample surface. To ensure consistency during probe update, a vacuum slice is introduced between the defocus plane and the sample in the initialization stage. By optimizing the thickness of this vacuum slice, the precise focal plane can be determined, which in turn allows the reconstruction of the top surface morphology. This approach is performed perfectly both in single-slice and multi-slice EP. In SSEP, where the sample can be seen as weak phase object and only contains 1 slice. Structure at a specific is often labeled as $O_r$, and electron probe is labeled as $P_r$. Then the exit-wave function can be written as:

$$\psi_{1,r} = \mathcal{P}\{P_r, \Delta d_r\} * O_r$$

Where $\Delta v_r$ is representing the vacuum layer thickness in specific scan position $r$. $\mathcal{P}\{W, z\}$ is the propagation operator, meaning the wave function $W$ propagating in free space with distance $z$. That can be written as:

$$\mathcal{P}\{W, z\} = \mathcal{F}^{-1}\{\mathcal{F}\{W\} * p(\boldsymbol{q}, z)\}$$

$\mathcal{F}$ represents Fourier transform. Then the exit-wave function can be shown as:

$$\psi_{1,r} = \mathcal{F}^{-1}\{\mathcal{F}\{P_r\} * p(\boldsymbol{q}, \Delta d_r)\} * O_r$$

Where Fresnel propagation factor $p(\boldsymbol{q}, \Delta d_r)$ is formed as:

$$p(\boldsymbol{q}, \Delta d_r) = \exp\left(-i\pi\lambda \cdot |\boldsymbol{q}|^2 \cdot \Delta d_r\right)$$

$\lambda$ is the wavelength; $\boldsymbol{q} = \sqrt{q_x^2 + q_y^2}$, where $q_x, q_y$ are reciprocal space coordinates. To accurately solve $\Delta d_r$, it is necessary to continuously compute its gradient during the iteration process and update the optimization. Its derivative can be calculated as:

$$\mathcal{g}_E(\Delta d_r) = \frac{\partial E}{\partial \Delta v_r} = 2 * (\chi_{1,r} - \psi_{1,r}) * \left(-\frac{\partial \psi_{1,r}}{\partial \Delta v_r}\right) = \mathcal{F}^{-1}\left\{\mathcal{F}\{P_r\} * -\left(\frac{\partial p(\boldsymbol{q}, \Delta d_r)}{\partial \Delta d_r}\right)\right\} * O_r$$

Its update step size can be determined using the least squares method as:

$$\mathcal{u}_E(\Delta v_r) = \frac{1}{2}\Re\{\frac{(\chi_{1,r} - \psi_{1,r}) * \left(\mathcal{g}_E(\Delta d_r)\right)^*}{|\mathcal{g}_E(\Delta d_r)|^2}\}$$

Based on this formula, sample's thickness variation can be calculated as well by changing the parameter from vacuum layer thickness $\Delta d_r$ to sample slice thickness $\Delta z_r$ in MSEP reconstruction. Here, one can choose to update the single-layer thickness and apply it to the entire stack or update each layer's thickness individually for more accurate results. The specific differentiation method is consistent with that described in previous work(40). After obtaining the derivative, it can be substituted back into the equation to determine the precise update step size. For the optimization and correction of real-valued parameters such as sample thickness, tilt angle, beam defocus, and beam scan position, the update step obtained from the least squares method based on gradient magnitude can be directly applied to update these real values after appropriate scaling. To ensure iterative convergence, a threshold value $\sigma$ is needed to prevent algorithm divergence. When the value oscillates within a region, narrowing the threshold range can help it gradually converge. This can be summarized by the following equation:

$$\tau(u, \sigma) = \begin{cases} u, & \text{if } |u| < |\sigma| \\ \dfrac{u \cdot \sigma}{|u|}, & \text{if } |u| > |\sigma| \end{cases}$$

**Simulation results of SAEP**

The simulated sample model is chosen to Beta zeolite. To better approximate experimental conditions, the simulation was performed using an electron dose of 1000 e$^-$/Å$^2$, a convergence semi-angle of 16 mrad, an accelerating voltage of 300 kV and a scan step size of 1.1 Å. To simulate the surface morphology and thickness variation of it, the defocus values were set from -80 nm to -120 nm in -5 nm increments, resulting in nine defocus planes for scanning, and the variable sample thicknesses from 12.7 nm (10 unit-cells) to 63.3 nm (50 unit-cells) in 5 unit-cells increments are placed into these nine scanning regions at the same time. The data from these nine groups were then stitched together and reconstructed simultaneously.

**Fig. 2A** shows the projected potential of the simulated model along the [010] direction. The atomic structure reconstructed via SAEP and conventional MSEP are shown in **Fig. 2B, C** respectively. **Fig. 2D** presents the defocus variation of the electron beam across different regions of the sample surface, which is used to simulate surface undulations. The defocus value map estimated by SAEP is displayed in (**Fig. 2E**). The surface morphology residual error map (**Fig. 2F**) shows even under low electron dose of 1000 e$^-$/Å$^2$, the maximum reconstruction error in each probe position remains below 10 nm. The position averaged error is at 0.83 nm. The ground truth, estimated value and residual error of thickness variation are shown in **Fig. 2G, H, I,** respectively. The error is also within 10 nm. To further investigate the source of the error, a series of extended simulations that cover broader aspects and discuss defocus estimation and thickness estimation separately, are presented in the supplementary text. These include the influence of convergence angle and electron dose on surface morphology reconstruction, thickness variation and error analysis for both surface and thickness estimation. In addition to that, the effect of regularization factor widely used during multi-slice reconstruction is discussed as well. The results demonstrate that the reconstruction error of surface morphology is independent of the depth of focus changes induced by varying the convergence angle; specifically, no significant difference is observed in the reconstruction error between 8 mrad and 24 mrad. In contrast, thickness estimation is more sensitive to the depth of focus, with higher depth resolution leading to more accurate thickness measurements. Additionally, increasing the regularization strength to obtain more converged reconstruction results significantly improves the precision of thickness estimation. At the case shown here, where the surface morphology and thickness variation are existed at the same time. The error originates from the low depth resolution caused by low electron dose and small convergence angle, which leads to inaccuracies in sample thickness estimation and, consequently, indirect errors in the estimation of surface morphology. Overall, the maximum error between the two is around 10 nm, while the average error across regions is approximately 5 nm, which is consistent with the depth resolution of MSEP reconstruction at a convergence angle of 16 mrad.

**Application of SAEP in real experiments**

**Sample preparation**

Beta zeolite was selected as the experimental sample. Its characteristic disordered structure induces extensive stacking faults within the particles, which in turn indirectly lead to abundant surface undulations and thickness variations at the macroscopic scale. The sample used for characterization was provided by our collaborators.

**Dose reduction and data collection**

The four-dimensional dataset is obtained by JEOL-GrandARM300 with MerlinEM detector, operated at 300 kV. A window of 256*257 pixels is scanned with a step size is of 1.123 Å. Semi-angle is chosen as 16 mrad. An overlap of 96% areas between adjacent scanning points is achieved with an under-focus of around 100 nm. To minimize the electron beam damage, low-dose condition was employed. The crystal was firstly tilted to zone axis under TEM mode using SAED, and then imaged under STEM mode. The electron probe of STEM mode was aligned in an amorphous carbon area before scanning the specimen. Electron dose used in experiment is about 1200 $e^-/Å^2$, measured by calibrated MerlinEM electron counts.

**Three-dimensional polymorph changes of Beta zeolite observed via conventional MSEP**

The MSEP reconstruction is carried out in two steps. The first step involves a coarse pre-reconstruction with a large defocus value to roughly estimate the sample thickness range and determine the position of the upper surface. This pre-reconstruction is performed with a defocus value initialized at -130 nm, a preset sample thickness of 130 nm, and a slice thickness of 10 nm, resulting in a total of 13 slices. During this step, scan position calibration is conducted, and the corrected probe positions are stored. The reconstruction estimates the upper focal plane of the sample at approximately -100 nm, with a maximum thickness of around 100 nm. The second step, a refined reconstruction, is performed based on this estimation.

To achieve better thickness sampling while ensuring an oversampling rate of at least 2 for the total signal, the second-step reconstruction is set with a defocus value of -100 nm, a thickness of 92 nm, and a slice thickness of 1 nm, yielding a total of 92 slices. To minimize incoherent signal interference, two states of probes are employed, with the scan positions directly utilizing the optimized positions from the preprocessing step. After reconstruction, post-processing is applied for inter-slice alignment.

The phase summation result of 92 reconstructed slices is shown in **Fig. 3A**. Benefiting from the 27 × 27 nm² field of view and the high density of stacking faults in Beta, phase transformations can be observed both within the same slice and across different slices (**Fig. 3B**) in the region marked by the light green box in **Fig. 3A**. As shown in **Fig. 3B**, in a single slice 41, a transition from polymorph-A (*ABAB*) to polymorph-B (*ABCA*) is observed. This kind of in-slice transition is caused by atomic dislocation such as split 12 ring (12R) marked with light green box in **Fig. 3B**. In addition to two-dimensional analysis, a transformation sequence of "polymorph-A to polymorph-B to polymorph-A" from the three-dimensional point of view is also identified between slices 41, 60, and 81.

The MSEP reconstruction results provide abundant atomic-level structural information for analysis. To investigate the stacking fault density in Beta zeolite, a machine learning-based classification was applied. The reconstructed results were categorized into three types according to their local structures: the first type corresponds to straight channels with no signal inside the 12R channel; the second type corresponds to blocked channels with signal detected inside the 12R. Signals originating from carbon films or vacuum are grouped as a third class (**Fig. 3C**). After classifying all 92 slices, the first 10 slices with insufficient channel information were excluded. This study utilized the *Trainable Weka Segmentation* plugin in ImageJ for machine-learning-based image recognition, classification and segmentation. To select an appropriate machine learning algorithm and verify segmentation accuracy, straight channels and blocked channels of the 80$^{th}$ slice, which has the highest resolution among all slices, were manually selected firstly, followed by training with different algorithm combinations. The training results were then compared with the manually segmented result using structural similarity index metric (SSIM), and the combination with the highest similarity was chosen for full-slices image segmentation. The final confirmed feature learning function set included *Gaussian blur, derivatives, structure*, and *difference of Gaussian*. The classifier used was *Fast Random Forest*. Details verifying the accuracy of classification can be found in supplementary text.

**Fig. 3D** shows the area fraction of straight channels $A_S$ and blocked channels $A_B$ in each slice. Both the area fractions and their ratio $A_S/A_B$ exhibit an abrupt change around slice 70, where $A_S$ reaches a minimum (45.2%) and $A_B$ reaches a maximum (36.1%) (**Fig. 3E**). This structural variation is clearly visualized by comparing the channel classification results in slices 60, 70, and 80 (**Fig. 3F**). Compared with the blocked channels observed in slices 60 and 80, slice 70 exhibits a large contiguous area of blocked channels, indicating the occurrence of a large-scale structural transformation within the zeolite particle. This structural transition is attributed to the internal stress release caused by stacking fault accumulation in Beta zeolite.

Geometric phase analysis (GPA) was further conducted at the location of an obvious fracture in slice 80 (highlighted by a purple box). In **Fig. 3G**, the results of overlaying three images are presented: the reconstructed phase, ideal structure model and stress mapping along $xx$ direction $\epsilon_{xx}$, to visually demonstrate the atomic displacement and the generation of stress. The results show that in slice 60, there is almost no deformation in this region, and the atomic structure is well aligned with the ideal model, with only occasional polymorphic stacking induced by overlapped12R. In slice 70, significant compressive stress was detected, accompanied by initial signs of cracking. Specifically, atoms on the left side of the fracture were shifted by 0.41 Å to the right relative to the ideal model, while atoms on the right side were shifted by 0.66 Å to the left. In slice 80, the fracture had fully developed; atoms on the left side were displaced by 0.55 Å to the right, and atoms on the right side were displaced by 0.66 Å to the left compared to the atomic model.

**Revealing the surface morphology and thickness variation of Beta zeolite by SAEP**

After the conventional MSEP reconstruction, the surface morphology and thickness variations of Beta can be intuitively observed by applying SAEP in a second reconstruction. (**Fig. 4A**). The surface morphology difference is approximately 30 nm, which aligns with the conventional MSEP reconstruction results (**Fig. 4B**). The thickness variation of the crystal is about 40 nm, primarily due to the elevated region at the upper right corner. The main body of the crystal has a thickness of approximately 60 nm, with some small regions around 50 nm or lower, caused by thickness irregularities due to structural loss in certain areas. As shown in **Fig. 4C**, the black lines outline regions where the crystal structure gradually disappears around slice 45, and then reappears around slice 55.

The lower surface morphology can be obtained by adding the upper surface morphology and sample thickness. According to the mapping, the lower surface of the crystal appears relatively smooth, with an overall undulation of less than 20 nm. Atomic-scale reconstruction results (**Fig. 4D**) confirm this conclusion. The lower surface undulation can be seen as the green-line outlined region gradually disappears, while the red-line outlined region still retains strong atomic signals.

**Building the relationship between atomic structure and thickness variation**

After obtaining atomic-level structural imaging through MSEP and macroscopic contour information through SAEP, establishing a connection between the two becomes a natural step. **Fig. 5A** shows the distribution of thickness regions, where it can be observed that the projected area of the 60–80 nm range accounts for 74.6% of the projected area of the 50–90 nm range, indicating that the main body of the crystal grain is within the 60–80 nm thickness range.

The proportion of straight channels within each thickness region is not the same. As the thickness increases, the proportion of straight channels decreases from 84.0% in the 50–60 nm region to 48.1% in the 80–90 nm region. **Fig. 5C** displays the mapping of channel transformation frequencies at different positions, revealing that as the sample thickness increases, the observed number of channel transformations gradually increases (from an average of 3.73 transformations in the 50–60 nm region to an average of 5.14 transformations in the 80–90 nm region). However, the average transformation probability per 10 nm thickness remains relatively constant, ranging from 60% to 70%. Although massive transformations cannot be counted for low depth resolution, the averaged transformation probabilities keep the same value.

Briefly summary, the proportion of straight channels over blocked channels in the Beta zeolite particle decreases as the increasing of thickness, but the transformation probability between straight and blocked channels remains almost identical. This suggests that blocked channels are more favorable for growth and thickening, whereas regions with straight channels are much more likely to be thinner.

**Application of SAEP in extremely low dose condition and large field of view imaging**

SAEP also has a significant impact on improving the quality of the reconstruction results at extremely low dose condition (<60 $e^-/Å^2$). **Fig. 6A** shows the conventional SSEP reconstruction result for an octahedron tip of MOF-MIL-101(Cr) at a dose level of 44 $e^-/Å^2$. SAEP also enables the observation of unique features across different regions of the sample in ultra large field of view characterization, such as local protrusions and large-area collapse in zeolites after irradiation damage. The atomic structure and surface morphology of an STW zeolite over a 100 × 100 $nm^2$ field of view are shown in **Fig. 6C** and **Fig. 6D**, respectively. Irradiation-damaged regions labeled as i, ii, and iii are marked with white (**Fig. 6C**) and green (**Fig. 6D**) boxes. In regions i and ii, localized damage results in surface protrusions and depressions, whereas region iii, which experienced large-area irradiation damage, exhibits structural collapse with an estimated depth of ~15 nm.

**Conclusion:**

In summary, this work leverages the information-rich nature of 4D-STEM single-scan data to successfully apply and further develop the MSEP algorithm. This enabled the simultaneous reconstruction of the local three-dimensional atomic framework of Beta zeolites and the characterization of surface morphology and thickness variations in the same region. By applying machine learning to classify the channel structures, a direct correlation was established between the microscopic pore configuration and the macroscopic thickness fluctuations. It was found that straight channels are more prevalent in thinner regions of the sample, while the probability of conversion between straight channels and blocked channels showed no clear dependence on thickness. Furthermore, the applicability of surface morphology characterization was explored under both extremely low-dose conditions and large fields of view, offering valuable insights for future in situ studies of zeolite structural evolution across different length scales.

**Acknowledgments**

E. Zhang thanks Y. Li for offering samples and Y. Lu, X. Wu, X. Hu, Y. Chen for dose calibrating.

**Funding:** This work is supported by:

National Natural Science Foundation of China (12335020)

Major State Basic Research Development Program of China (2022YFA1603703)

Strategic Priority Research Program of the Chinese Academy of Sciences (No. XDB 37040303)

**Author contributions:**

This article is supervised under Y. Ma and H. Jiang. E. Zhang finished the algorithms, simulations, data analysis and wrote the article. Z. Dong collected the 4D-data. X. Han built the atomic model. J. Zhang. gave advises to article written. All authors discussed the results and gave vital advises on the manuscript.

**Competing interests:** Authors declare that they have no competing interests.

**Data and materials availability:** All datasets are available on Zenodo xxx.

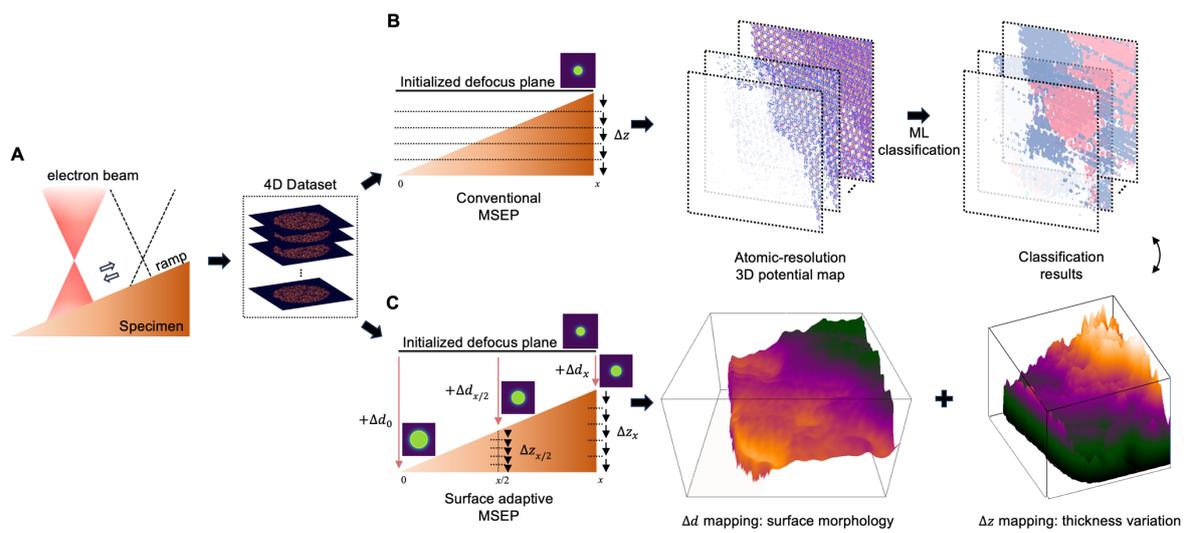

**Fig. 1 Schematic of linking microscopic atomic structures with macroscopic morphology via MSEP and surface adaptive MSEP reconstructions.** (A) 4D-STEM data collection of a specimen with uniform surface and thickness. (B) Implementing conventional MSEP reconstruction to obtain high-resolution atomic structure for further classification and segmentation. (C) Implementing surface adaptive MSEP to obtain surface morphology and thickness variation mapping.

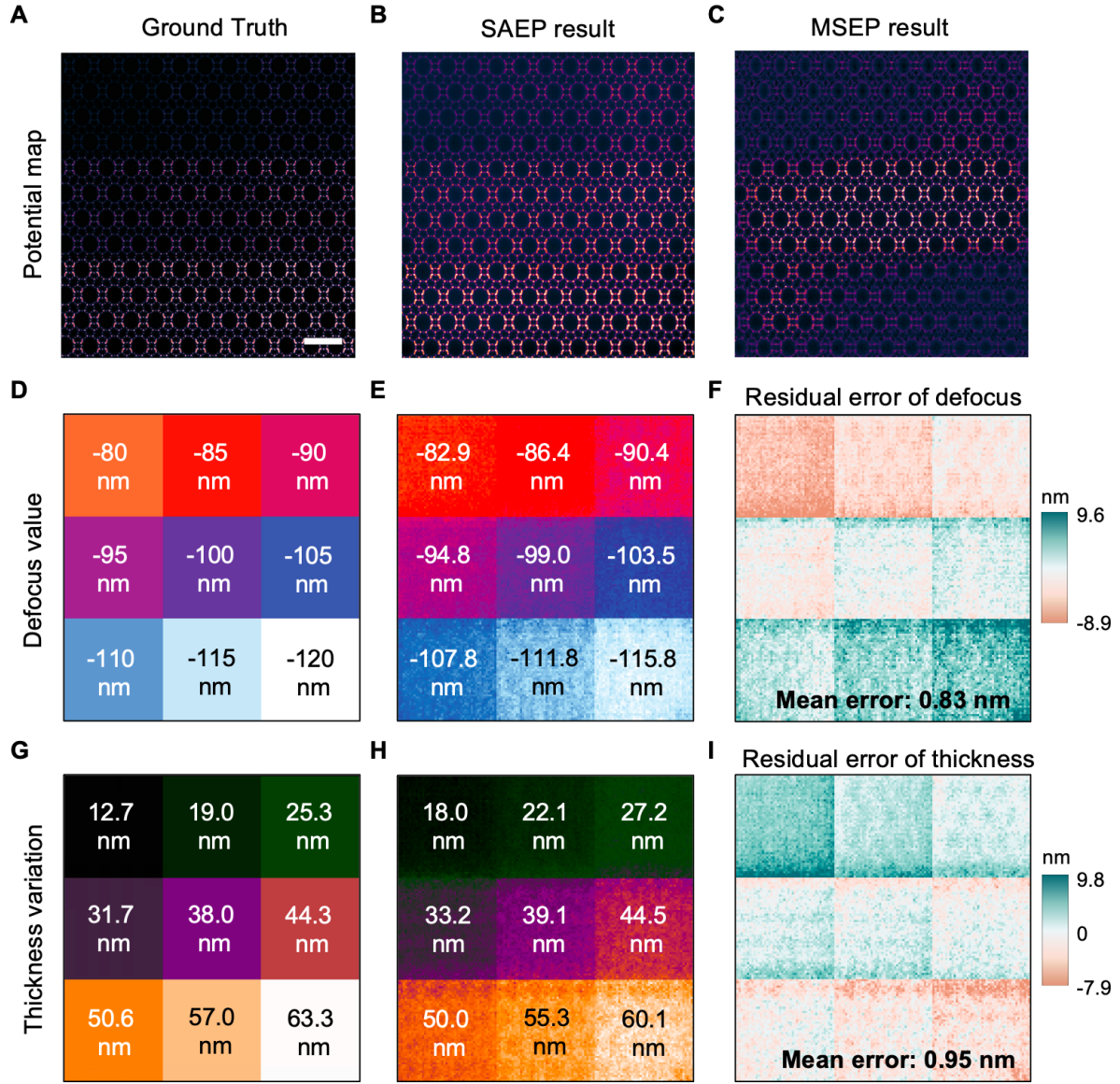

**Fig. 2. Surface morphology reconstruction of simulated Beta zeolite by SAEP under 1000 e⁻/Å².** (A) Simulated potential map of variable thickness from 12.7 nm to 60.4 nm Beta zeolite along [010] direction. Reconstructed potential map with (B) and without (C) SAEP algorithm. (D) The ground truth and estimated defocus value via SAEP. (F) The residual error between (D)(E). (G)(H) The ground truth and estimated thickness value via SAEP. (I) The residual error between (G)(H). Scale bar: 2 nm.

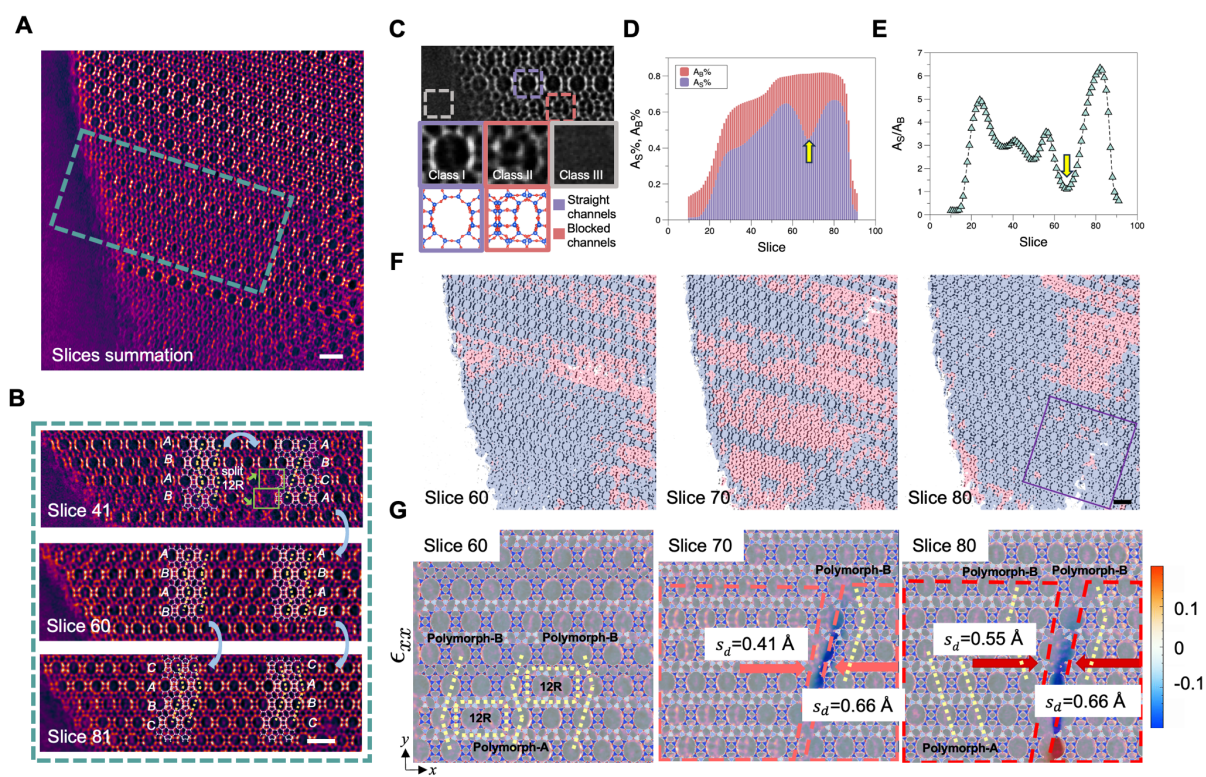

**Fig. 3. Large-scale structural transformation and stress relaxation in Beta.** (A) Phase summation of 92 reconstructed slices. (B) Polymorph transformations observed within slice 41 and between slices 41, 60, and 81. Scale bar: 2 nm. (C) MSEP reconstruction results categorized into three structural types. (D) The area fractions of straight and blocked channels in each slice, along with the area ratio of straight to blocked channels in (E). The yellow arrows in (D)(E) indicate a large-scale structural transformation at this location. (F) Classification of channels at the structural transformation site in the upper (slice 60), middle (slice 70), and lower (slice 80) sections. (G) Overlayed results of structure model (polyhedral), reconstructed phase image and strain mapping along $xx$ direction for the fractured crystal structure in slice 80, highlighted by the purple box. Scale bar: 2 nm.

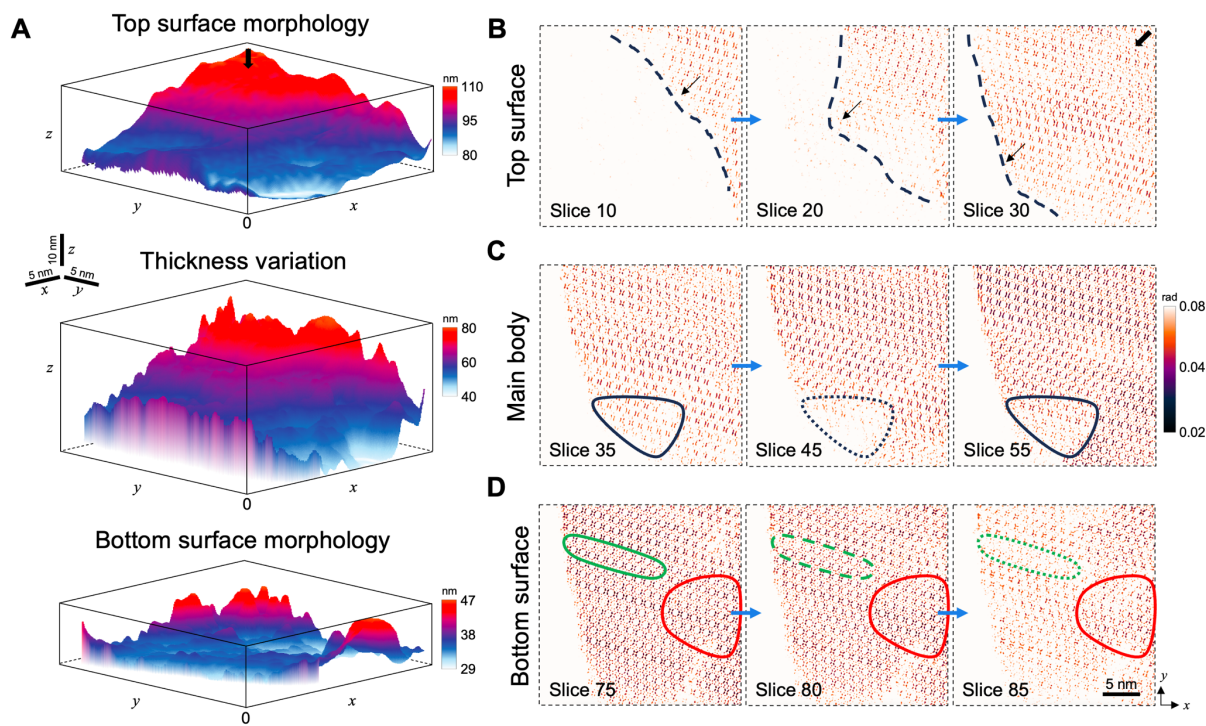

**Fig. 4. The surface-adaptive EP method is used to determine the upper and lower surface morphology and thickness variations of Beta crystals within the reconstruction field of view.** (A) shows the surface undulations of the upper and lower surfaces of the Beta particle, along with the sample thickness variation. The surface undulation range is 30 nm, which aligns with the reconstruction results shown in (B), The consistent viewing angles are marked with black bold arrows in (A) and (B) and the electron beam focus is taken as the zero point of height. (C) The black lines mark areas of structural loss in the thinner regions. (D) The lower surface is generally smooth, with variations of less than 20 nm. However, atomic-scale structural analysis reveals regional inhomogeneity, such as the gradual disappearance of the green region, while the red region consistently retains stronger atomic signals.

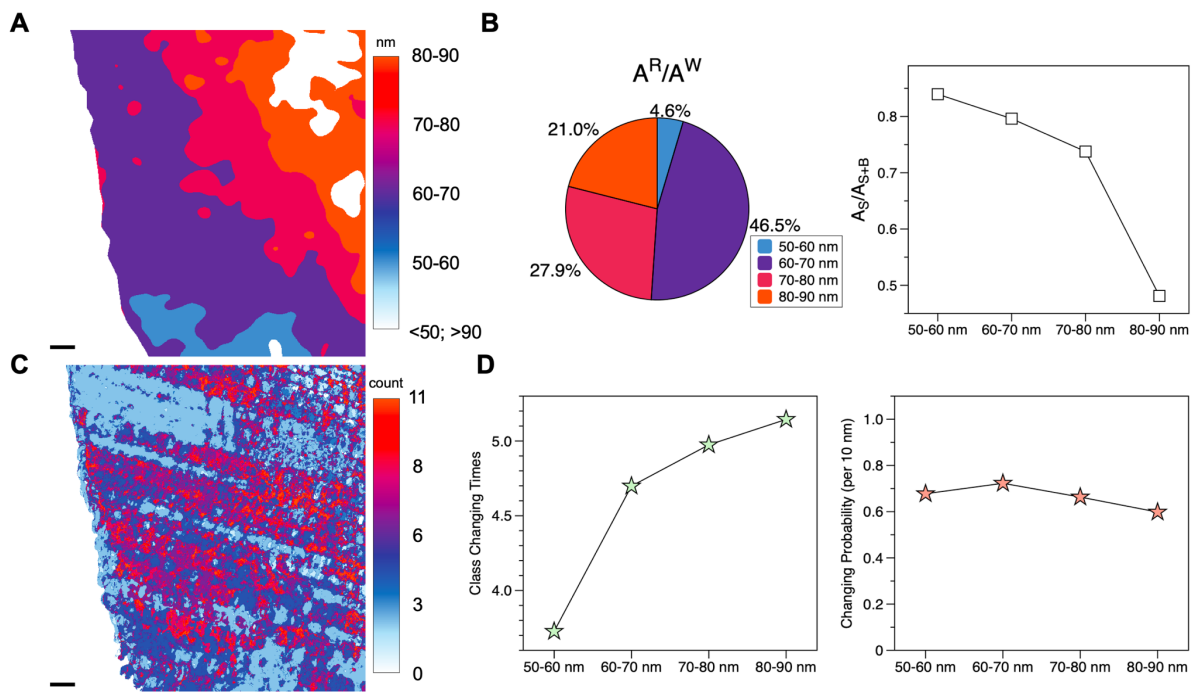

**Fig. 5. Correlation between sample thickness and structural transformation.** (A) Thickness map of the Beta crystal grain. (B) The area fraction $A^R$ of each thickness region relative to the total projected area $A^W$, and the area fraction of straight channels $A_S$ within each thickness region relative to the total channel area $A_{S+B}$. (C) Mapping of the transformation frequency between straight and blocked channels. (D) The average number of transformations between the two types of channels in different regions, as well as the average probability of transformation occurring. Scale bar: 2 nm.

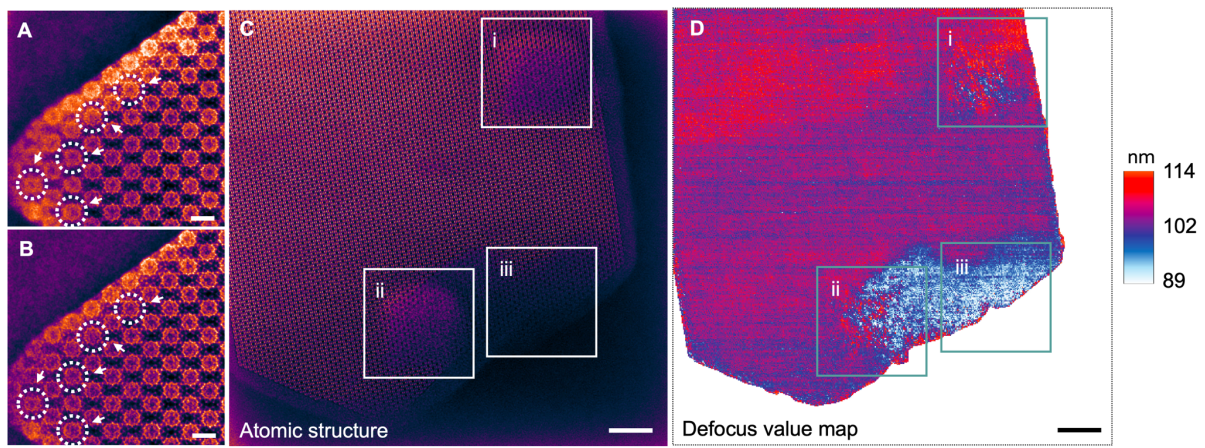

**Fig. 6. Application of SAEP in low-dose and large-FOV imaging. Comparison of** reconstruction results without (A) and with (B) surface-adaptive EP at dose level of 44 e$^-$/Å$^2$. The atomic structures at the top and edges of the sample become clearly visible after the SAEP reconstruction (areas marked by white circles and arrows). Atomic structure (C) and defocus value mapping (D) of STW zeolite within a FOV of 100*100 nm$^2$. i, ii and iii are three regions that show different morphology type under beam radiation damages. Scale bar in (A, B): 5 nm, scale bar in (C)(D): 10 nm.